\documentclass[review]{elsarticle}

\usepackage{lineno,hyperref}
\usepackage{amssymb,amsmath,amsthm}
\usepackage{subcaption,caption}
\usepackage{appendix}
\usepackage{color}
\allowdisplaybreaks

\modulolinenumbers[5]

\journal{Journal of Mathematical Analysis and Applications}

\begin{document}

\begin{frontmatter}

\title{Hysteresis bifurcation and application to delayed Fitzhugh-Nagumo neural systems\tnoteref{mytitlenote}}
\tnotetext[mytitlenote]{This work was supported by the Natural Sciences and Engineering Research Council of Canada.}

\author{L. Chen}
\ead{L477chen@uwaterloo.ca}

\author{S. A. Campbell\corref{mycorrespondingauthor}}
\cortext[mycorrespondingauthor]{Corresponding author}
\ead{sacampbell@uwaterloo.ca}

\address{Department of Applied Mathematics, University of Waterloo, Waterloo, ON, N2L 3G1, Canada}

\begin{abstract}
Hysteresis dynamics has been described in a vast number of biological experimental studies. Many such studies are phenomenological and a mathematical appreciation has not attracted enough attention.  In the paper, we explore the nature of hysteresis and study it from the dynamical system point of view by using the bifurcation and perturbation theories. We firstly make a classification of hysteresis according to the system behaviours transiting between different types of attractors. Then, we focus on a mathematically amenable situation where hysteretic movements between the equilibrium point and the limit cycle are initiated by a subcritical Hopf bifurcation and a saddle-node bifurcation of limit cycles. We present a analytical framework by using the method of multiple scales to obtain the normal form up to the fifth order. Theoretical results are compared with time domain simulations and numerical continuation, showing good agreement. Although we consider the time-delayed FitzHugh-Nagumo neural system in the paper, the generalization should be clear to other systems or parameters. The general framework we present in the paper can be naturally extended to the notion of bursting activity in neuroscience where hysteresis is a dominant mechanism to generate bursting oscillations. 
\end{abstract}


\end{frontmatter}


\section{Introduction}
Hysteresis widely exists in biology from microscopic cell biology \cite{Angeli:2004} , genetics \cite{Andrews:2013} and neuroscience  \cite{Noori:2011} up to macroscopic bio-mechanical properties of organs such as the eye \cite{DelM:2011} and muscle \cite{RAMOS:2017}. More examples can be found in ecological and epidemic models, such as the spruce budworm model \cite{Ludwig:1978}, coral reef model \cite{Blackwood:2010} and savanna and forest model \cite{Staver:2012}. In particular, hysteresis is one essential mechanism to generate bursting oscillations which play important roles in communication between neurons \cite{Izhikevich:2000}. However, there had been few mathematical investigations of this biological process  until the discovery of a number of molecular mechanisms with bistable dynamical behavior by the early 1990s \cite{NOORI:2014}. In addition, influenced by mathematical treatments to physical and engineering systems, most studies in biology concentrate on identification and modelling by inserting hysteresis operators into mathematical equations, e.g. the Preisach model of ATP hysteresis \cite{Noori:2011} and the models for bacteria growth or prey-predator systems \cite{Kopf:2006}. However, there do exist a variety of biological models without explicitly embedded hysteresis operators, but still distinctly demonstrating hysteresis, e.g. systems introduced in \cite{ Angeli:2004,Gardner:2000,NOY:1975,Kuzn:1995}. In addition, hysteresis is not new and has been widely observed in a variety of disciplines, such as  material science, mechanics, electronics and economics. As a result of years of interdisciplinary work, the definitions of hysteresis are useful but different in specific contexts. A stringently mathematical and universal definition has not yet appeared.

\begin{figure}[ht]
\centering
\includegraphics[width=0.5\textwidth]{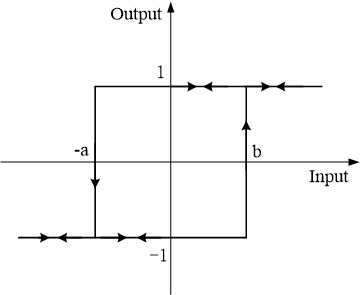}
\caption{Binary hysteresis (also called relay) with output $\in \{-1,1\}$ and width $a+b$.}
\label{fig:relay}
\end{figure}

Three essential components are usually used to characterize hysteresis: lagging, rate-independence and looping behaviour \cite{MORRIS:2012}. These can be easily understood from a simple input/output plot of hysteresis shown in Fig.~\ref{fig:relay}. Lagging means that the output lags the input; rate-independence indicates that the output only depends on the values, not the rates of change, of the input; and the looping behaviour implies that the output is affected by the previous values of the input, demonstrating a memory effect. Although all the three components are generally regarded as crucial features of hysteresis, contradictory examples are not uncommon. Therefore, we need to understand the nature of hysteresis.

Recently, a new definition was proposed from the dynamical system point of view.

\textbf{Definition \cite{MORRIS:2012}.} \emph{A hysteretic system is one which has (1) multiple stable equilibrium points and (2) dynamics that are considerably faster than the time scale at which inputs are varied.}

The definition points out two main features of a dynamical system with the property of hysteresis. One is multistability, another is dramatic changes with respect to the slower input. Further, it implies that hysteresis by nature can be understood by analysis of the multistability displayed in the bifurcation diagram where dramatic transitions occur between multistable attractors by varying the relatively constant bifurcation parameters.  From this perspective, hysteresis dynamics has a strong link to the notion of bursting oscillations. Bursting oscillations, as an important neural activity, are usually studied via bifurcation theory and analysis of fast-slow systems, where the slow variables are treated as parameters of the fast dynamics \cite{Izhikevich:2000,Izhikevich:2007}. In some examples, the fast subsystem exhibits multistability, which leads to  a hysteretic loop visiting alternately one of two different attractors corresponding to resting and spiking states, respectively.

\begin{figure}[htbp]
\centering
\includegraphics[width=0.3\textwidth]{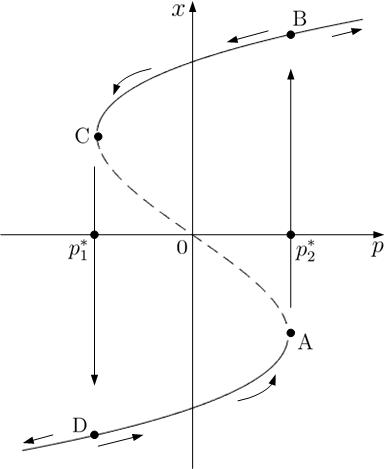}
\caption{Hysteresis generated by two saddle-node bifurcations in the equation $\dot x = p+x-x^3$. Modified from \cite{NOORI:2014}.}
\label{fig:hysteresis-saddle-node}
\end{figure}

Moreover, in bifurcation theory terminology, hysteresis dynamics above has an equivalent name, \emph{hysteresis bifurcation} which is a type of reversible catastrophe. Catastrophic bifurcation occurs  when a microscopic variation of a parameter triggers a macroscopic movement from one attractor to another. If the system can be driven back to the initial attractor, the catastrophe is called reversible. Fig.~\ref{fig:hysteresis-saddle-node} depicts one kind of hysteresis bifurcation induced by two saddle-node bifurcations at critical points $p_1^*$ and $p_2^*$, respectively. Between $p_1^*$ and $p_2^*$, the system is bistable with two stable equilibrium points. As the bifurcation parameter $p$ increases, the trajectory of the system slowly slides up along the lower stable path (the solid curve from $D$) until it reaches the right knee,  $A$. At this moment, it quickly jumps to point $B$, another attractor leading to a higher stable branch. This jump is "considerably faster than the time scale at which" the bifurcation parameter is varied. Likewise, the backwards procedure goes down along the upper stable branch. Upon reaching the left knee, $C$, the system jumps to the lower branch and slide left. If the parameter $p$ is varied back and forth, the trajectory of the system follow closely the loop $A \rightarrow B \rightarrow C \rightarrow D$ resulting in a reversible catastrophe. The loop is called hysteretic loop, or briefly, hysteresis, which is analogous to Fig.~\ref{fig:relay} with the three characteristics: lagging, rate-independence and looping behaviour. Generally, attractors in the hysteretic loop could be a stable equilibrium point, a stable periodic orbit, an attractive torus, even a strange attractor. Therefore, we suggest replacing the phrase "equilibrium points" in the definition with a more general expression, "attractors".

Hysteresis is easy to be understood conceptually, but some of the attributes are quite difficult to study mathematically \cite{Izhikevich:2007}. Our work aims to mathematically investigate hysteresis from the dynamical system point of view. Bifurcation and perturbation theories are used to analytically study qualitative or topological changes of the trajectories of the nonlinear dynamics. We start with classification of hysteresis initiated from all possible codimension-one bifurcations of equilibrium points. Then, we perform a specific analysis on the time-delayed FitzHugh-Nagumo neural system to show how the subcritical Hopf bifurcation and saddle-node bifurcation of limit cycles generate hysteresis. It may be the simplest instance to form a hysteretic loop transiting between equilibrium points and limit cycles. The method of multiple timescales is used to derive a normal form up to the fifth order. While we focus on a specific system, the methods we use can be applied to any model involving ordinary or delay differential equations.

The paper is organized as follows. In Sec. 2, we summarize the possible situations where hysteresis bifurcation occurs and classify them. Section 3 presents the theoretical framework for hysteresis analysis of the time-delayed FitzHugh-Nagumo neuron.  In Sec. 4, we validate our analytical results against  solutions obtained with the time domain simulation and numerical continuation. Finally, we conclude our findings in Sec. 5.

\section{Classification of hysteresis bifurcation}
A bifurcation indicates a transition from one qualitative type of dynamics to another \cite{GH:1983,Kuzn:1998}. Thus, we classify a hysteresis bifurcation by its generation mechanism, that is, what kinds of attractors are involved in this transition.

\subsection{Transition between equilibrium points}\label{sec:HysteresisClassification1}
The neural system has two classic types of attractors: the resting state (quiescence) and periodic spiking. These two states correspond to a stable equilibrium point and a limit cycle attractor, respectively. Switching between two stable resting states has been observed in many experiments, e.g. \cite{Tasaki:1959}. Hysteresis formed by transitions between equilibrium points can also be seen in the Hodgkin-Huxley model for the squid axon where the transmembrane voltage is the bifurcation variable and the external potassium concentration acts as the bifurcation parameter \cite{Aihara:1983}.

\begin{figure}[ht]
\centering
\includegraphics[width=0.4\textwidth]{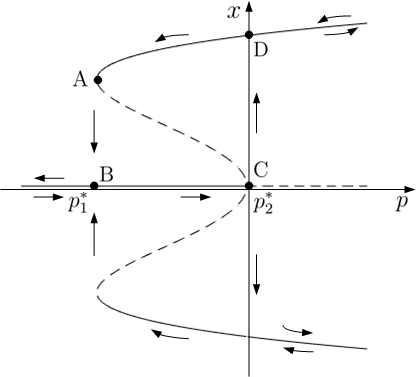}
\caption{Two hysteresis bifurcations generated by two saddle-node bifurcations at $p_1^*$ and a subcritical pitchfork bifurcation at $p_2^*$ in the equation $\dot x = px + x^3 - x^5$.}
\label{fig:hysteresis-saddle-pitchfork}
\end{figure}

\begin{figure}[htbp]
\centering
\includegraphics[width=0.5\textwidth]{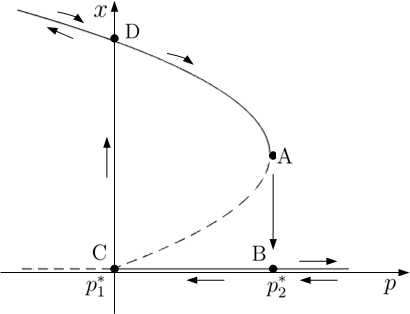}
\caption{Hysteresis generated by a saddle-node bifurcation at $p_2^*$ and a transcritical bifurcation at $p_1^*$ in the equation $\dot x = -px + 4x^2- x^3$.}
\label{fig:hysteresis-saddle-transcritical}
\end{figure}

Three codimension-one bifurcations involve equilibrium points in a dynamical system: saddle-node bifurcation, transcritical bifurcation and pitchfork bifurcation. Hysteretic loops generated by the three bifurcations are summarized in Fig. \ref{fig:hysteresis-saddle-node}, \ref{fig:hysteresis-saddle-pitchfork} and \ref{fig:hysteresis-saddle-transcritical}. Mathematically each example can be described by a one-dimensional nonlinear equation with the bifurcation parameter $p$.

The bifurcation diagram in Fig.~\ref{fig:hysteresis-saddle-node} is generated from a dynamical system expressed as
\begin{equation*}
  \dot x = p + x - x^3,
\end{equation*}
where $\dot x = \mathrm{d}x/\mathrm{d}t$ is the derivative of the variable $x$ with respect to time $t$. From $p + x - x^3=0$ (equilibrium condition) we find $p = x^3 - x$. By using the extreme value theory, letting $dp/dx = 0$, one derives the mirrored critical values at $p_1^*=-\frac{2\sqrt{3}}{9}$ and $p_2^*=\frac{2\sqrt{3}}{9}$. Then, through bifurcation analysis, we know that the system has two saddle-node bifurcations at $p_1^*$ and $p_2^*$, respectively, with zero eigenvalues at equilibria $A$ ($x=-1/\sqrt{3}$) and $C$ ($x = 1/\sqrt{3}$).  The hysteretic loop  has a width of
\begin{equation*}
  \chi = p_2^*-p_1^* =  \frac{2\sqrt{3}}{9} - (-\frac{2\sqrt{3}}{9}) = \frac{4\sqrt{3}}{9}.
\end{equation*}

Consider the nonlinear equation
\begin{equation*}
  \dot x = px + x^3 - x^5.
\end{equation*}
The bifurcation diagram in Fig.~\ref{fig:hysteresis-saddle-pitchfork} shows multi-stability and hysteresis of this dynamical system. By bifurcation analysis one can derive that two saddle-node bifurcations occur at $p_1^* = -1/4$ and a subcritical pitchfork bifurcation at $p_2^*=0$.  Two symmetric hysteretic loops are generated with the range calculated as 
\begin{equation*}
  \chi = p_2^*-p_1^* = 0 - (-1/4) =  \frac{1}{4}.
\end{equation*}

In addition, variables of biological systems mostly are positive, which can lead to different bifurcations. The hysteretic curve of Fig.~\ref{fig:hysteresis-saddle-transcritical} looks like the flipped copy of the upper part of Fig.~\ref{fig:hysteresis-saddle-pitchfork}. However, the hysteresis generation mechanism is not the same.  Let us consider the following nonlinear equation with $x > 0$ for physical reasons,
\begin{equation*}
  \dot x = -px + 4x^2 - x^3.
\end{equation*}
By bifurcation analysis, we can see that a transcritical bifurcation at $p_1^* = 0$ and a saddle-node bifurcation at $p_2^* = 4$ complete the hysteretic loop in Fig.~\ref{fig:hysteresis-saddle-transcritical} with a  width of 
\begin{equation*}
  \chi = p_2^*-p_1^* = 4-0 = 4.
\end{equation*}
A similar bifurcation diagram, except shifting  to the right some units, can be found in the exploited population model \cite{NOY:1975}.

\begin{figure}[htbp]
    \centering
    \includegraphics[width=0.55\textwidth]{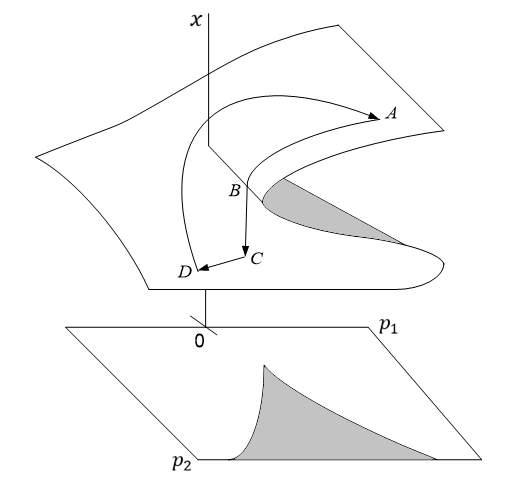}
    \caption{Cusp bifurcation of the equation $\dot x = p_1 + p_2x - x^3$. Modified from \cite{DERCOLE:2011}.}
    \label{fig:cusp}
\end{figure}

Besides codimension-one bifurcations, the cusp catastrophe, a codimension-two bifurcation, can give rise to hysteresis. Fig.~\ref{fig:cusp} depicts a two-parameter bifurcation diagram of cusp from the equation
\begin{equation*}
    \dot x = p_1 + p_2x - x^3.
\end{equation*}
Within the cusp-shaped grey region illustrated in the $(p_1,p_2)$ parameter plane, there are three equilibrium points present. Outside of this region, there is only one equilibrium point. 
Compared with two macroscopic jumps in Fig.~\ref{fig:hysteresis-saddle-node}, \ref{fig:hysteresis-saddle-pitchfork} and \ref{fig:hysteresis-saddle-transcritical}, the hysteretic loop here is formed by a smooth movement along the arrow $C \rightarrow D \rightarrow A \rightarrow B$ and a catastrophic transition from $B$ to $C$.

\subsection{Transition between equilibrium points and limit cycles}\label{sec-hys-generation-hopf}
From the examples above  we can see that the saddle-node bifurcation frequently appears in forming a hysteretic loop. Thus, it should not be surprising that the counterpart saddle-node bifurcation of limit cycles can also be involved in hysteresis. Hysteresis involving movement between an equilibrium point and a limit cycle has  been found experimentally in the squid axon and numerically in the Hodgkin-Huxley model in response to the variation of the injected  bias current \cite{Guttman:1980}. This has been explained by the combination of a subcritical Hopf bifurcation and a saddle-node bifurcation of limit cycles which initiate hysteretic dynamics in the model.

\begin{figure}[htbp]
    \centering
    \includegraphics[width=0.6\textwidth]{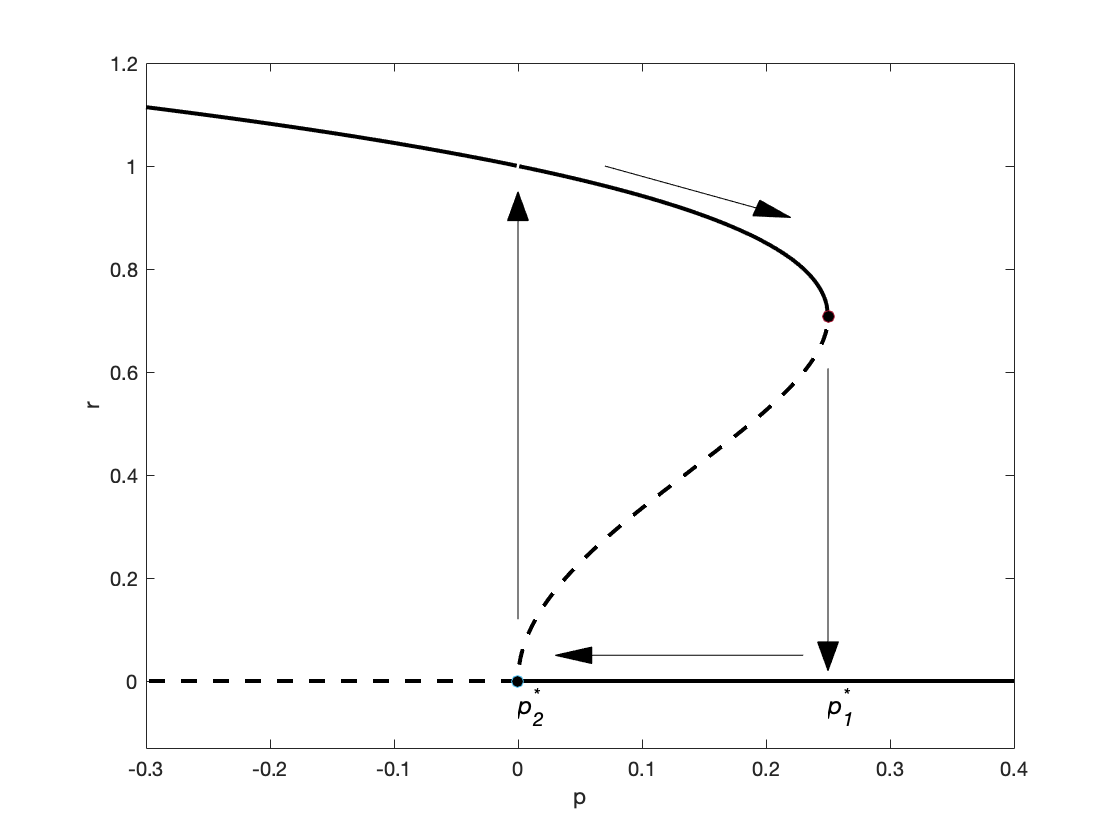}
    \caption{Sketch of the bifurcation diagram in the equation $\dot r = - pr + r^3-r^5 $. Hysteresis initiated by a saddle-node bifurcation of limit cycles at $p^*_1$ and a subcritical Hopf bifurcations at $p^*_2$. The arrows show one possible movement. Solid (dash) lines correspond to stable (unstable) solutions.}
    \label{fig:hysteresis_5th}
\end{figure}

A mathematical understanding of such a hysteresis bifurcation can be achieved by reducing the system model to a fifth order normal form with the equation of the amplitude of periodic orbits,
\begin{equation}\label{eq-SLE51a}
  \dot r    =  \alpha_r r + \beta_r r^3 + c_r r^5, 
\end{equation}
where $\alpha_r$, $\beta_r$ and $c_r$ are real values \cite{GH:1983,Kuzn:1998}. The solutions of (\ref{eq-SLE51a}) are
\begin{equation}\label{eq-SLE5-solution}
    r_1=0, \quad r_{2,3}=\sqrt{\frac{-\beta_r \pm \sqrt{\big (\beta_r)\big)^2-4\alpha_r c_r}}{2c_r}},
\end{equation}
where $r_1 = 0$ corresponds to the equilibrium point, and the periodic orbit exists when either $r_{2,3}$ or both have positive real values. The stability of the solutions is evaluated by the sign of the Jacobian
\begin{equation}\label{eq-SLE5-Jacobi}
    J = \alpha_r + 3\beta_r r^2 + 5c_r r^4.
\end{equation}
Fig.~\ref{fig:hysteresis_5th} illustrates a sketch of bifurcation diagram of (\ref{eq-SLE51a}). The system undergoes a subcritical Hopf bifurcation at the critical point $p^*_2$, where $\alpha_r(p_2^*) = 0$, and a saddle-node bifurcation of limit cycles at $p^*_1$, where the local extremum of $\alpha_r(p)$ with respect to $r$ reaches, that is, $\alpha_r(p_1^*) = \beta_r^2/(4c_r)$. When $p > p^*_1$, the system has only one stable equilibrium; when $p < p^*_2$, the system has an unstable equilibrium point and a stable limit cycle; a bistable region appears between $p^*_1$ and $p^*_2$, where the system trajectory transits between a stable equilibrium point and a stable limit cycle. In the next section, we will show how to derive the normal form (\ref{eq-SLE51a}) and investigate such a hysteresis bifurcation by application to a time-delayed neural model. The relevant methods can be generalized to other situations.

Hysteresis may also occur due to a sequence of bifurcations that occurs in a particular model. For example, \cite{DERCOLE:2011} introduces a more complex hysteresis found in a tritrophic food chain model. Here, catastrophic transitions between the equilibrium point and the prey-predator limit cycle are initiated by a transcritical bifurcation and a homoclinic bifurcation.

\subsection{Transition between limit cycles}
We have seen that hysteresis may result from the coexistence of two stable equilibrium points, it is also possible that two stable limit cycles coexist. The possible corresponding behaviours in a neural system are spikes fired with different periods. For example, it has been shown that bistability between in-phase and anti-phase oscillations can occur in models for systems of two coupled neurons \cite{Campbell:2012, Kunec:2001, Park:2016, Ryu:2020}. This has been linked to pitchfork bifurcations of limit cycles \cite{Campbell:2012, Park:2016} and subcritical Hopf bifurcations \cite{Ryu:2020}. Similar phenomena have been found in many biological systems, including an ionic model of ventricular membrane, where hysteretic transitions between periodic orbits with respective 1:1 and 2:1 rhythms occurs at different driving frequencies \cite{Yehia:1999}. Mathematically, one possibility to generate such a hysteresis bifurcation can be achieved by two saddle-node bifurcations of limit cycles, similar to Fig.~\ref{fig:hysteresis-saddle-node}.

\section{Hysteresis of the time-delayed FitzHugh-Nagumo neurons}
In the section, we develop a theoretical analysis of hysteresis induced from a subcritical Hopf bifurcation and a saddle-node bifurcation of limit cycles by application to a time-delayed FitzHugh-Nagumo (FHN) neural system.

The FHN model \cite{Fitzhugh:1961} is a two-dimensional simplification of the Hodgkin-Huxley equations describing spike generation. Although not clearly derivable from biology, the model has becomes a central model in mathematical neuroscience and is simple enough to allow analytical developments. Further, the influence of synaptic delays on system dynamics cannot be ignored or underestimated in the field of neuroscience \cite{Ghosh:2008}. This has motivated many time-delayed neuron models, including the one proposed in \cite{PLANT:1981}.

\subsection{Time-delayed FitzHugh-Nagumo neurons}\label{sec:FhN-model}
The time-delayed FHN model introduced in \cite{PLANT:1981} is modelled by a system of delay differential equations,
\begin{equation}\label{eq-FN-timedelay}
  \begin{split}
    \dot v & = v(t) - \frac{1}{3}v^3(t)-w(t)+\mu\big(v(t-\tau)-v_0\big), \\
    \dot w & = \rho\big(v(t)+a-bw(t)\big),
  \end{split}
\end{equation}
where $\rho$ represents the timescale ratio between the membrane potential $v$ and the recovery variable $w$, the time delay $\tau>0$ and $\mu$ is the strength of the feedback, positive for excitatory and negative for inhibitory feedback.

For $\mu =0$, the system (\ref{eq-FN-timedelay}) has an equilibrium point at $(v_0,w_0)$ given by
\begin{equation}\label{eq-FN-point}
  \begin{split}
    0 & = v_0-\frac{1}{3}v_0^3-\frac{1}{b}(v_0+a), \\
    w_0 & =(v_0+a)/b.
  \end{split}
\end{equation}
Moreover, under the following conditions
\begin{equation}\label{eq-FN-para}
  0< \rho <1, \quad 0< b <1, \quad 1-2b/3 < a < 1,
\end{equation}
and
\begin{equation*}
    1-b\rho < v_0^2 < 1+b\rho+2\sqrt{\rho},
\end{equation*}
the equilibrium is unique and a stable focus \cite{Fitzhugh:1969}. Define $x=v-v_0$, $y=w-w_0$ and the vector $\boldsymbol{u}= [x, y]^T$ ($'T'$ means transpose), the equilibrium point is transformed to zero in the transformed model:
\begin{equation}\label{eq-FN-timedelay2-matrix}
  \dot{\boldsymbol{u}}   = A\boldsymbol{u}(t) + \mu B\boldsymbol{u}(t-\tau) + \boldsymbol{f}\big(\boldsymbol{u}(t)\big),
\end{equation}
where
\begin{equation*}
  A  = \left(
        \begin{array}{ll}
          1-v_0^2, & -1 \\
          \rho,    & -\rho b \\
        \end{array}
      \right), \; B  = \left(
        \begin{array}{ll}
          1 & 0\\
          0 & 0\\
        \end{array}
      \right),  \; \boldsymbol{f}\big(\boldsymbol{u}(t)\big)  =\left(
         \begin{array}{c}
           -v_0x^2(t)-\frac{1}{3}x^3(t) \\
           0 \\
         \end{array}
       \right).
\end{equation*}

\subsection{Normal form of hysteresis bifurcation}\label{sec:MMS}
In the section, we demonstrate the hysteresis bifurcation structure of (\ref{eq-FN-timedelay2-matrix}). The method of multiple scales \cite{DAS:2002,Johnson:2005} is used to obtain the normal form by expanding the evolution of the dynamical system (\ref{eq-FN-timedelay2-matrix}) around the Hopf location.

Let us take $\mu$ as the bifurcation parameter and define
\begin{equation}\label{eq-bifur-mu}
  \mu = \mu_c + \varepsilon^2\delta_2,
\end{equation}
where $\mu_c$ is the Hopf bifurcation point, $0<\varepsilon \ll 1$ is a small quantity that quantifies the magnitude of the oscillations close to $\mu_c$ and  $\delta_2$ takes the values $\pm 1$ depending on the side of the Hopf point. Then, we seek a three-timescale five-order expansion of the solution of (\ref{eq-FN-timedelay2-matrix}) in the neighborhood of $\mu = \mu_c$ in the form
\begin{equation}\label{eq-MMS-u1}
  \boldsymbol{u}(t,\varepsilon) = \sum_{k=1}^5{\varepsilon^k \boldsymbol{U}_k(T_0, T_2, T_4)}=\sum_{k=1}^5{\varepsilon^k \left(
        \begin{array}{c}
            X_k(T_0, T_2, T_4) \\
            Y_k(T_0, T_2, T_4) 
        \end{array}
        \right)}.
\end{equation}
Here, $T_0 = t$ is the fast timescale, $T_2 = \varepsilon^2 t$ and $T_4 = \varepsilon^4 t$ are the first and second slow timescales, respectively. The derivative with respect to $t$ is transformed into
\begin{equation}\label{eq-MMS-dt1}
  \frac{d}{dt}=\frac{\partial }{\partial T_0}  + \varepsilon^2 \frac{\partial }{\partial T_2} + \varepsilon^4 \frac{\partial }{\partial T_4} 
\end{equation}
Given by $\boldsymbol{u} = [x,y]^T$ and (\ref{eq-MMS-u1}), $\boldsymbol{f}\big(\boldsymbol{u}(t)\big)$ in (\ref{eq-FN-timedelay2-matrix}) is rewritten as
\begin{equation}\label{eq-MMS-nonlinear}
    \begin{split}
      \boldsymbol{f}\big(\boldsymbol{u}(t)\big) & =  \varepsilon^2 \left(
            \begin{array}{c}
                -v_0X_1^2 \\
                0 \\
            \end{array}\right) + \varepsilon^3 \left(
            \begin{array}{c}
                -2v_0X_1X_2-\frac{1}{3}X_1^3 \\
                0 \\
            \end{array}\right) \\
            & \hspace{1cm} + \varepsilon^4 \left(
            \begin{array}{c}
                -v_0X_2^2-2v_0X_1X_3-X_1^2X_2 \\
                0 \\
            \end{array}\right)  \\
           & \hspace{1cm} + \varepsilon^5 \left(
            \begin{array}{c}
                -2v_0X_2X_3-2v_0X_1X_4-X_1^2X_3-X_2^2X_1 \\
                0 \\
            \end{array}\right)\\
         & \equiv  \sum_{k\geq 2}{\varepsilon^k \boldsymbol{f}_k(\boldsymbol{U}_1, \boldsymbol{U}_2, \boldsymbol{U}_3, \boldsymbol{U}_4, \boldsymbol{U}_5)}
    \end{split}
  \end{equation}
In addition, the delay term $\boldsymbol{u}(t-\tau)$ in (\ref{eq-FN-timedelay2-matrix}) is expressed in terms of the scales $T_0$, $T_2$ and $T_4$ as  
\begin{multline}\label{eq-MMS-u-delay}
    \boldsymbol{u}(t-\tau,\varepsilon)     
    =  
    \varepsilon \boldsymbol{U}_{1\tau}
    +
    \varepsilon^2\boldsymbol{U}_{2\tau}
    +
    \varepsilon^3
    \left (
    \boldsymbol{U}_{3\tau}-
    \tau 
    \frac{\partial{\boldsymbol{U}_{1\tau}}}
    {\partial{T_2}}
    \right ) \\
    + 
    \varepsilon^4
    \left (
    \boldsymbol{U}_{4\tau}-
    \tau 
    \frac{\partial{\boldsymbol{U}_{2\tau}}}
    {\partial{T_2}}
    \right )
    +
    \varepsilon^5
    \left (
    \boldsymbol{U}_{5\tau}-
    \tau 
    \frac{\partial{\boldsymbol{U}_{1\tau}}}
    {\partial{T_4}}
    -\tau 
    \frac{\partial{\boldsymbol{U}_{3\tau}}}
    {\partial{T_2}}
    \right )
\end{multline}
where $\boldsymbol{U}_{i\tau}=\boldsymbol{U}_i(T_0-\tau, T_2, T_4), i = 1, 2, 3$. By substituting (\ref{eq-bifur-mu})-(\ref{eq-MMS-u-delay}) into (\ref{eq-FN-timedelay2-matrix}) and matching these terms by their $\varepsilon$ order, we obtain five differential equations as follows:
\begin{equation}\label{eq-FN-MMS1}
  \frac{\partial \boldsymbol{U}_1}{\partial T_0}-A\boldsymbol{U}_1-\mu_cB\boldsymbol{U}_{1\tau}=0
\end{equation}
\begin{equation}\label{eq-FN-MMS2}
  \frac{\partial \boldsymbol{U}_2}{\partial T_0} - A\boldsymbol{U}_2 - \mu_cB\boldsymbol{U}_{2\tau} = \boldsymbol{f}_2
\end{equation}
\begin{equation}\label{eq-FN-MMS3}
  \frac{\partial \boldsymbol{U}_3}{\partial T_0} - A\boldsymbol{U}_3 - \mu_cB\boldsymbol{U}_{3\tau} = -\frac{\partial \boldsymbol{U}_1}{\partial T_2}-\tau\mu_cB\frac{\partial \boldsymbol{U}_{1\tau}}{\partial T_2} 
  + \delta_2B\boldsymbol{U}_{1\tau}+\boldsymbol{f}_3,
\end{equation}
\begin{equation}\label{eq-FN-MMS4}
  \frac{\partial \boldsymbol{U}_4}{\partial T_0} - A\boldsymbol{U}_4 - \mu_cB\boldsymbol{U}_{4\tau}  = -\frac{\partial \boldsymbol{U}_2}{\partial T_2}-\tau\mu_cB\frac{\partial \boldsymbol{U}_{2\tau}}{\partial T_2} +\delta_2B\boldsymbol{U}_{2\tau}+\boldsymbol{f}_4
\end{equation}
\begin{multline}\label{eq-FN-MMS5}
  \frac{\partial \boldsymbol{U}_5}{\partial T_0} - A\boldsymbol{U}_5 - \mu_cB\boldsymbol{U}_{5\tau} = -\frac{\partial \boldsymbol{U}_1}{\partial T_4}-\tau\mu_cB\frac{\partial \boldsymbol{U}_{1\tau}}{\partial T_4} -\frac{\partial \boldsymbol{U}_3}{\partial T_2} \\
        - \tau\mu_cB\frac{\partial \boldsymbol{U}_{3\tau}}{\partial T_2}+ \delta_2B\boldsymbol{U}_{3\tau}-\delta_2B\tau\frac{\partial \boldsymbol{U}_{1\tau}}{\partial T_2} + \boldsymbol{f}_5
\end{multline}

\subsubsection{Derivation of the third-order normal form}
Next, we will solve the foregoing three equations (\ref{eq-FN-MMS1})-(\ref{eq-FN-MMS3}) one by one to derive the third-order normal form of the Hopf bifurcations. 

Solving (\ref{eq-FN-MMS1}) is a typical nonlinear eigenvalue problem that has a general solution, 
\begin{equation}\label{eq-FN-MMS-u1}
  \boldsymbol{U}_1 = W(T_2,T_4)\boldsymbol{q}e^{iw_cT_0} + c.c.
\end{equation}
Here, $c.c.$ stands for the complex conjugate of the preceding terms and has the form of $\overline{W}(T_2,T_4)\overline{\boldsymbol{q}}e^{-iw_cT_0}$; $W(T_2,T_4)$ is the complex amplitude depending on the slow timescales, $T_2$ and $T_4$ will be determined in later steps; $s = iw_c$ is the eigenvalue at the Hopf point $\mu=\mu_c$ where the system is marginally stable and can be obtained by solving the characteristic equation ${\rm det}(M_s) =0$ with
\begin{equation}\label{eq-FN-MMS-eigenvalue}
  M_s \equiv sI - A - \mu_cBe^{-s\tau}.
\end{equation}
$\boldsymbol{q}$ is the corresponding eigenvector, which is not unique. Here, we use a general notation to the 2-D eigenvector,
\begin{equation}\label{eq-FN-MMS-vector1}
  \boldsymbol{q} =  \left (
                      \begin{array}{c}
                        X_1^W \\
                        Y_1^W \\
                      \end{array}
                    \right ).
\end{equation}
$\boldsymbol{q}$ can be also taken to be 
\begin{equation}\label{eq-FN-MMS-vector}
  \boldsymbol{q} = \left (
                      \begin{array}{c}
                        1 \\
                        \frac{\rho}{\rho b+iw_c} \\
                      \end{array}
                    \right ),
\end{equation} 
for the specific FHN system (\ref{eq-FN-timedelay2-matrix}).

Substituting (\ref{eq-FN-MMS-u1}) into (\ref{eq-FN-MMS2}) yields
\begin{equation}\label{eq-FN-MMS2-1}
  \frac{\partial \boldsymbol{U}_2}{\partial T_0} - A\boldsymbol{U}_2 - \mu_cB\boldsymbol{U}_{2\tau} = |W|^2\boldsymbol{F}_2^{|W|^2} + \big( W^2\boldsymbol{F}_2^{W^2}e^{2iw_cT_0} + c.c. \big),
\end{equation}
where
\begin{subequations}\label{eq-FN-MMS2-2}
 \begin{align}
   \boldsymbol{F}_2^{|W|^2}  & = \left(
                        \begin{array}{c}
                          -2v_0|X_1^W|^2 \\
                          0 \\
                        \end{array}
                      \right)  = -2v_0|X_1^W|^2 \left(\begin{array}{c}
                          1 \\
                          0 \\
                        \end{array}\right)\\
   \boldsymbol{F}_2^{W^2}   & = \left(
                          \begin{array}{c}
                            -v_0(X_1^W)^2 \\
                            0 \\
                          \end{array}
                        \right)  = -v_0(X_1^W)^2 \left(\begin{array}{c}
                          1 \\
                          0 \\
                        \end{array} \right).
 \end{align}
\end{subequations}
When using (\ref{eq-FN-MMS-vector}), then gives
\begin{equation}\label{eq-FN-MMS2-21}
   \boldsymbol{F}_2^{|W|^2}   =  -2v_0 \left(\begin{array}{c}
                          1 \\
                          0 \\
                        \end{array}\right), \quad
   \boldsymbol{F}_2^{W^2}    =  -v_0 \left(\begin{array}{c}
                          1 \\
                          0 \\
                        \end{array} \right).
\end{equation}
Here, the superscripts of $\boldsymbol{F}$ show the dependence on the amplitude $W$ and subscripts indicate the corresponding equations (\ref{eq-FN-MMS1})-(\ref{eq-FN-MMS5}). The notation is borrowed from \cite{ORCHINI:2016} and will be employed throughout the paper.

Assume $\boldsymbol{U}_2$ has the same form as the forcing term in (\ref{eq-FN-MMS2-1})
\begin{equation}\label{eq-FN-MMS-u2}
  \boldsymbol{U}_2 = |W|^2\boldsymbol{U}_2^{|W|^2} + \big( W^2\boldsymbol{U}_2^{W^2}e^{2iw_cT_0} + c.c. \big).
\end{equation}
Substituting it into (\ref{eq-FN-MMS2-1}), taking the Laplace transform with respect to $T_0$ and matching the terms according to their amplitude dependence yield specific expressions (see details in the appendix).

Next, substituting (\ref{eq-FN-MMS-u1}) and (\ref{eq-FN-MMS-u2}) into the differential equation (\ref{eq-FN-MMS3}), we have 
\begin{multline}\label{eq-FN-MMS31}
  \frac{\partial \boldsymbol{U}_3}{\partial T_0} -  A\boldsymbol{U}_3 - \mu_cB\boldsymbol{U}_{3\tau} = \left ( -\frac{\partial W}{\partial T_2}H\boldsymbol{q} + W\boldsymbol{F}_3^W \right.\\ 
  \left. + |W|^2W\boldsymbol{F}_3^{|W|^2W}  \right )e^{iw_cT_0} 
          + W^3\boldsymbol{F}_3^{W^3}e^{3iw_cT_0} + c.c.
\end{multline}
where
\begin{equation}\label{eq-FN-MMS-P}
  H = I + \tau\mu_cB e^{-iw_c\tau}.
\end{equation} 
To guarantee (\ref{eq-FN-MMS31}) has solutions, a solvability condition has to be satisfied. The condition is that the sum of the resonant forcing terms, that is, the terms with $e^{iw_cT_0}$ on the right-hand side of (\ref{eq-FN-MMS31}), should be orthogonal to every solution of the adjoint homogeneous problem \cite{ODEN:2010}. In this case, the adjoint problem is
\begin{equation*}
  M_{iw_c}^\dagger \boldsymbol{q}^\dagger = 0,
\end{equation*}
where $M_{iw_c}^\dagger$ is the Hermitian of the matrix $M_{iw_c}$ and has the form
\begin{equation}\label{eq-FN-MMS3-adjoint-matrix}
  M_{iw_c}^\dagger \equiv -iw_cI - A^T - \mu_cB^Te^{iw_c\tau}.
\end{equation}
Taking the inner product of the resonant forcing terms of (\ref{eq-FN-MMS31}) with $\boldsymbol{q}^\dagger$ yields the solvability condition,
\begin{equation}\label{eq-FN-MMS3-solve-condition}
  \Big \langle \boldsymbol{q}^\dagger, -\frac{\partial W}{\partial T_2}H\boldsymbol{q} + W\boldsymbol{F}_3^W + |W|^2W\boldsymbol{F}_3^{|W|^2W} \Big \rangle = 0,
\end{equation}
which is rewritten as 
\begin{equation}\label{eq-FN-SLE3}
  \frac{\partial W}{\partial T_2} = \alpha_3W +\beta_3|W|^2W.
\end{equation}
Here, the complex values $\alpha_3$ and $\beta_3$ are calculated as
\begin{equation}\label{eq-FN-SLE3-coefficient}
  \alpha_3 = \frac{\langle \boldsymbol{q}^\dagger, \boldsymbol{F}_3^W \rangle}{\langle \boldsymbol{q}^\dagger, H\boldsymbol{q}\rangle}, \quad \beta_3 = \frac{\langle \boldsymbol{q}^\dagger, \boldsymbol{F}_3^{|W|^2W} \rangle}{\langle \boldsymbol{q}^\dagger, H\boldsymbol{q}\rangle}.
\end{equation}
For easy of calculation, we can choose a unique $\boldsymbol{q}^\dagger$ by imposing the following condition, 
\begin{equation*}
    \langle \boldsymbol{q}^\dagger, \boldsymbol{q} \rangle = \langle \boldsymbol{q}, \boldsymbol{q}^\dagger \rangle = \bar{\boldsymbol{q}}^T\boldsymbol{q}^\dagger = 1.
\end{equation*}

In addition, the solution of (\ref{eq-FN-MMS31}) has the form
\begin{equation}\label{eq-FN-MMS-u3}
  \boldsymbol{U}_3 = W\boldsymbol{U}_3^We^{iw_cT_0} + |W|^2W\boldsymbol{U}_3^{|W|^2W}e^{iw_cT_0} + W^3\boldsymbol{U}_3^{W^3}e^{3iw_cT_0} + c.c.
\end{equation}
The appendix gives the specific expressions.

Eq. (\ref{eq-FN-SLE3}) is a third-order normal form usually used to understand the Hopf bifurcation \cite{GH:1983,Kuzn:1998}. Substituting $W(t) = r(t)e^{i\theta(t)}$ into (\ref{eq-FN-SLE3}) and taking real and imaginary parts of the resulting equation, we get the expressions in polar coordinates as
\begin{subequations}\label{eq-SLE31}
  \begin{align}
  \dot r      & =   {\rm Re}(\alpha_3)r + {\rm Re}(\beta_3)r^3, \label{eq-SLE31a} \\
  \dot \theta & =   {\rm Im}(\alpha_3)  + {\rm Im}(\beta_3)r^2, \label{eq-SLE31b}
\end{align}
\end{subequations}
where $\alpha_3$ and $\beta_3$ are called the Landau coefficients. The amplitude equation (\ref{eq-SLE31a}) has solutions
\begin{equation}\label{eq-SLE3-solutions}
    r_1 = 0, \quad r_2 = \sqrt{-\frac{{\rm Re}(\alpha_3)}{{\rm Re}(\beta_3)}}.
\end{equation}
The stability of the solutions is determined by the sign of the eigenvalue $\lambda$ evaluated at the solutions. These are
\begin{equation}\label{eq-SLE3-eigenvalues}
    \lambda(r_1)={\rm Re}(\alpha_3), \quad \lambda(r_2)=-2{\rm Re}(\alpha_3)
\end{equation}

\begin{figure}[htbp]
    \centering
    \includegraphics[width=0.6\textwidth]{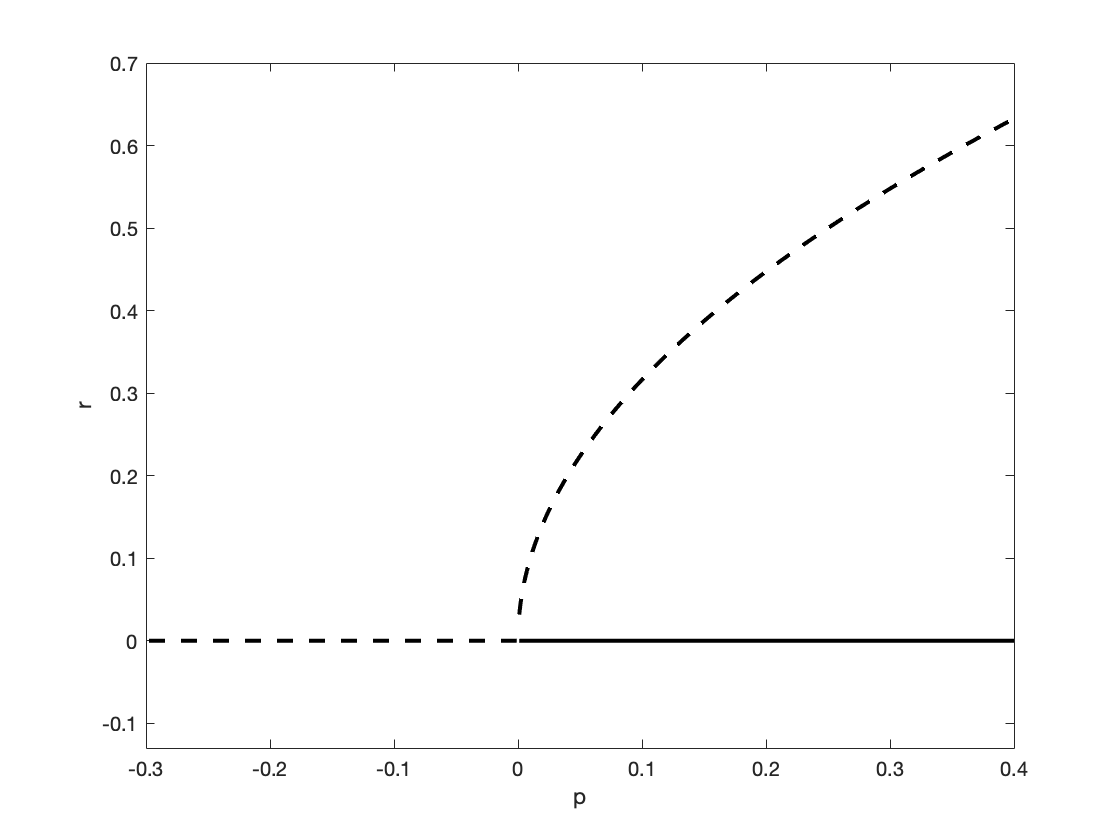}
    \caption{Sketch of the bifurcation diagram obtained from $\dot r =  r^3 - pr$.}
    \label{fig:hysteresis_3rd}
\end{figure}

Based on bifurcation theory, a subcritical Hopf bifurcation occurs when ${\rm Re}(\beta_3) > 0$ \cite{GH:1983,Kuzn:1998}. A sketch of the bifurcation diagram can be see in Fig.~\ref{fig:hysteresis_3rd}. Before the Hopf point, the system has a unique solution and it is an unstable equilibrium point; after that, the system has one stable equilibrium and one unstable limit cycle. As can be seen, no hysteresis is produced in the third-order normal form (\ref{eq-SLE31}). We need continue to solve the differential equations (\ref{eq-FN-MMS4}) and (\ref{eq-FN-MMS5}) to obtain more information.  

\subsubsection{Derivation of the fifth-order normal form}
Substituting (\ref{eq-FN-MMS-u1}),  (\ref{eq-FN-MMS-u2}) and (\ref{eq-FN-MMS-u3}) into the differential equation (\ref{eq-FN-MMS4}), we have 
\begin{multline}\label{eq-FN-MMS41}
  \frac{\partial \boldsymbol{U}_4}{\partial T_0} -  A\boldsymbol{U}_4  - \mu_cB\boldsymbol{U}_{4\tau}  = |W|^4\boldsymbol{F}_4^{|W|^4} + |W|^2\boldsymbol{F}_4^{|W|^2}+ \left(W^2\boldsymbol{F}_4^{W^2}e^{2iw_cT_0} \right. \\
       \left. + |W|^2W^2\boldsymbol{F}_4^{|W|^2W^2}e^{2iw_cT_0} + W^4\boldsymbol{F}_4^{W^4}e^{4iw_cT_0} + c.c.\right),
\end{multline}
One can see that there are no resonant terms on the right-hand side of the equation. Then, used the ansatz  
\begin{multline}\label{eq-FN-MMS-u4}
  \boldsymbol{U}_4 = |W|^4\boldsymbol{U}_4^{|W|^4} + |W|^2\boldsymbol{U}_4^{|W|^2} + \left( W^2\boldsymbol{U}_4^{W^2}e^{2iw_cT_0}  \right. \\
      \left. + |W|^2W^2\boldsymbol{U}_4^{|W|^2W^2}e^{2iw_cT_0} + W^4\boldsymbol{U}_4^{W^4}e^{4iw_cT_0} + c.c.   \right),
\end{multline}
a set of expressions is readily derived in the appendix.

Similarly, by substituting the previous solutions of $U_1$ to $U_4$, the differential equation (\ref{eq-FN-MMS5}) can be rewritten as 
\begin{multline}\label{eq-FN-MMS51}
  \frac{\partial \boldsymbol{U}_5}{\partial T_0} - A\boldsymbol{U}_5 - \mu_cB\boldsymbol{U}_5(T_0-\tau) = \left(-\frac{\partial W}{\partial T_4}H\boldsymbol{q} +  WF_5^W \right.\\
      \left. + |W|^2WF_5^{|W|^2W} + |W|^4WF_5^{|W|^4W} \right)e^{iw_cT_0} \\
      + \left( |W|^2W^3F_5^{|W|^2W^3} + W^3F_5^{W^3}\right)e^{3iw_cT_0} + W^5F_5^{W^5}e^{5iw_cT_0} + c.c.
\end{multline}
One can see that there are resonant terms with $e^{iw_cT_0}$ on the right-hand side of the equation. Therefore, by applying the solvability condition on the resonant terms as before, we obtain the fifth-order normal form at the timescale $T_4$,
\begin{equation}\label{eq-FN-SLE5}
  \frac{\partial W}{\partial T_4} = \alpha_5W +\beta_5|W|^2W + c_5|W|^4W,
\end{equation}
where the Landau coefficients are 
\begin{equation}\label{eq-FN-SLE5-coefficient}
  \alpha_5 = \frac{\langle \boldsymbol{q}^\dagger, \boldsymbol{F}_5^W \rangle}{\langle \boldsymbol{q}^\dagger, H\boldsymbol{q}\rangle}, \quad \beta_5 = \frac{\langle \boldsymbol{q}^\dagger, \boldsymbol{F}_5^{|W|^2W} \rangle}{\langle \boldsymbol{q}^\dagger, H\boldsymbol{q}\rangle}, \quad c_5 = \frac{\langle \boldsymbol{q}^\dagger, \boldsymbol{F}_5^{|W|^4W} \rangle}{\langle \boldsymbol{q}^\dagger, H\boldsymbol{q}\rangle}.
\end{equation}

Eventually, the final fifth-order normal form is derived by combining (\ref{eq-FN-SLE3}), (\ref{eq-FN-SLE5}) and using the scaling $T_4 = \varepsilon^2T_2$ \cite{ORCHINI:2016} as
\begin{equation}\label{eq-FN-SLE51}
\begin{split}
  \frac{{\rm d}W}{{\rm d}T_2} & = \frac{\partial W}{\partial T_2} + \frac{\partial W}{\partial T_4}\frac{\partial T_4}{\partial T_2} \\
                  & = \left( \alpha_3+\varepsilon^2\alpha_5\right)W + \left( \beta_3+\varepsilon^2\beta_5 \right)|W|^2W + \varepsilon^2c_5|W|^4W
\end{split}
\end{equation}
Substituting the polar representation $W = re^{i\theta}$, we have the normal form with the amplitude and phase of limit cycle solutions,
\begin{subequations}\label{eq-FN-SLE52}
  \begin{align}
  \frac{{\rm d}r}{{\rm d}T_2}        & =   {\rm Re}(\alpha)r + {\rm Re}(\beta)r^3 + {\rm Re}(c)r^5, \label{FN-SLE52a} \\
  \frac{{\rm d} \theta}{{\rm d} T_2} & =   {\rm Im}(\alpha)  + {\rm Im}(\beta)r^2 + {\rm Im}(c)r^4. \label{FN-SLE52b}
\end{align}
\end{subequations}
Here, $\alpha = \alpha_3 + \varepsilon^2\alpha_5$, $\beta = \beta_3 + \varepsilon^2\beta_5$ and $c = \varepsilon^2c_5$. One can see that (\ref{FN-SLE52a}) has the same expression as (\ref{eq-SLE51a}), where hysteretic transitions between a equilibrium point and a limit cycle is generated by a subcritical Hopf bifurcation and a saddle-node bifurcation of limit cycles, as shown in Fig.~\ref{fig:hysteresis_5th}. The solution of the phase equation (\ref{FN-SLE52b}) reads:
\begin{align}
    \theta 
    & = 
    \left ( 
    \mathrm{Im}(\alpha) 
    + \mathrm{Im}(\beta) r^2_{2,3} 
    + \mathrm{Im}(c) r^4_{2,3}
    \right) T_2 \nonumber\\
    & =
    \left ( 
    \mathrm{Im}(\alpha) 
    + \mathrm{Im}(\beta) r^2_{2,3} 
    + \mathrm{Im}(c) r^4_{2,3}
    \right) 
    \varepsilon^2 T_0 \nonumber\\
    & \equiv
    \Delta w_{2,3} T_0,
\end{align}
where $r_{2,3}$ are limit cycle solutions of (\ref{FN-SLE52a}). $\Delta w$ can be looked as the frequency shift between the fundamental oscillation frequency of limit cycles and the marginally stable frequency $w_c$. In addition, we use the ansatz
\begin{multline}
   \boldsymbol{U}_5 =  \left (|W|^4 W \boldsymbol{U}_5^{|W|^4 W} 
  + |W|^2 W \boldsymbol{U}_5^{|W|^2 W} 
  + W \boldsymbol{U}_5^W \right )e^{i w_c T_0}...
  \\
  + \left(|W|^2 W^3 \boldsymbol{U}_5^{|W|^2 W^3} 
  + W^3 \boldsymbol{U}_5^{W^3} \right) e^{3 i w_c T_0}
  + W^5 \boldsymbol{U}_5^{W^5} e^{5 i w_c T_0}
  + c.c.,
\end{multline} 
and substitute it into (\ref{eq-FN-MMS51}),  the solution of $U_5$ is readily obtained (see the appendix).

To derive the analytical solution of the delayed FHN model up to the fifth order, we combine the power expansion (\ref{eq-MMS-u1}), solutions at each order $\boldsymbol{U}_1$ to $\boldsymbol{U}_5$ and the solution of the normal form (\ref{eq-FN-SLE52}). The final expression reads 
\begin{align}
    \boldsymbol{u} & =  \varepsilon \boldsymbol{U}_1 
           + \varepsilon^2 \boldsymbol{U}_2 
           + \varepsilon^3 \boldsymbol{U}_3 
           + \varepsilon^4 \boldsymbol{U}_4 
           +\varepsilon^5 \boldsymbol{U}_5 
           +\mathcal{O}(\varepsilon^6)  
           \nonumber\\
      & = \varepsilon^2r^2 \boldsymbol{U}_2^{|W|^2}
          + \varepsilon^4 r^4 \boldsymbol{U}_4^{|W|^4}
          + \varepsilon^4 r^2 \boldsymbol{U}_4^{|W|^2}... \nonumber\\
      & \qquad    +\Big [
          \varepsilon r \boldsymbol{U}_1^W
          +\varepsilon^3 r \boldsymbol{U}_3^W
          +\varepsilon^3 r^3 \boldsymbol{U}_3^{|W|^2W}
          +\varepsilon^5 r^5 \boldsymbol{U}_5^{|W|^4W}... \nonumber \\
      & \qquad \qquad \qquad \qquad  \quad 
          +\varepsilon^5 r^3 \boldsymbol{U}_5^{|W|^2W}
          +\varepsilon^5 r \boldsymbol{U}_5^W
          \Big ] e^{i(w_c+\Delta w)T_0}
          ... \nonumber\\
      & \qquad + \left [
          \varepsilon^2 r^2 \boldsymbol{U}_2^{W^2}
          + \varepsilon^4 r^4 \boldsymbol{U}_4^{|W|^2W^2}
          + \varepsilon^4 r^2 \boldsymbol{U}_4^{W^2}
          \right ]e^{2i(w_c+\Delta w)T_0}...
          \nonumber\\
      & \qquad + \left [    
          \varepsilon^3 r^3 \boldsymbol{U}_3^{W^3}
          + \varepsilon^5 r^5 \boldsymbol{U}_5^{|W|^2W^3}
          + \varepsilon^5 r^3 \boldsymbol{U}_5^{W^3}
          \right ]e^{3i(w_c+\Delta w)T_0}...
          \nonumber\\
      & \qquad + 
          \varepsilon^4 r^4 \boldsymbol{U}_4^{W^4} e^{4i(w_c+\Delta w)T_0}
          + \varepsilon^5 r^5 \boldsymbol{U}_5^{W^5} e^{5i(w_c+\Delta w)T_0}
      +c.c. 
      + \mathcal{O}(\varepsilon^6).
\end{align}

The generation mechanism of hysteresis is simpler when only involving equilibrium points. Complexity increases with the involvement of limit cycles. Here, we have focused on a relatively analytically tractable case and presented a analytical framework by applying the method of multiple scales to the delayed FHN neuron model. Our approach can be easily extended to other systems, or to other bifurcation parameters, such as $\tau$, to investigate the impact of time delays on dynamics.

\section{Numerical analysis}
Using the method of multiple scales to derive the normal form, even a low-order one, may be a lengthy and tedious process. However, such a procedure is standard and can be automatised with symbolic solvers. See for example \cite{SANCHEZ:1996}.

In this section, we show some numerical results to confirm the analytical expressions. The parameter values are chosen as $a = 0.7$ and $b = 0.8$, commonly used in the literature.  As for $\rho = 0.08$ and $\tau = 60$, originally used in \cite{PLANT:1981}, we have found by DDE-BIFTOOL \cite{Sieber:2017} that the system undergoes more complicated bifurcations, including not only the subcritical Hopf bifurcation and saddle-node bifurcation of limit cycles, but also the period-doubling bifurcation of limit cycles and torus bifurcation. Therefore, we set $\rho = 0.5$ and $\tau = 15$ to obtain the relatively simpler and illustrative bifurcation structure to show hysteretic dynamics. 

\begin{figure}[htbp]
    \centering
    \includegraphics[width=0.7\textwidth]{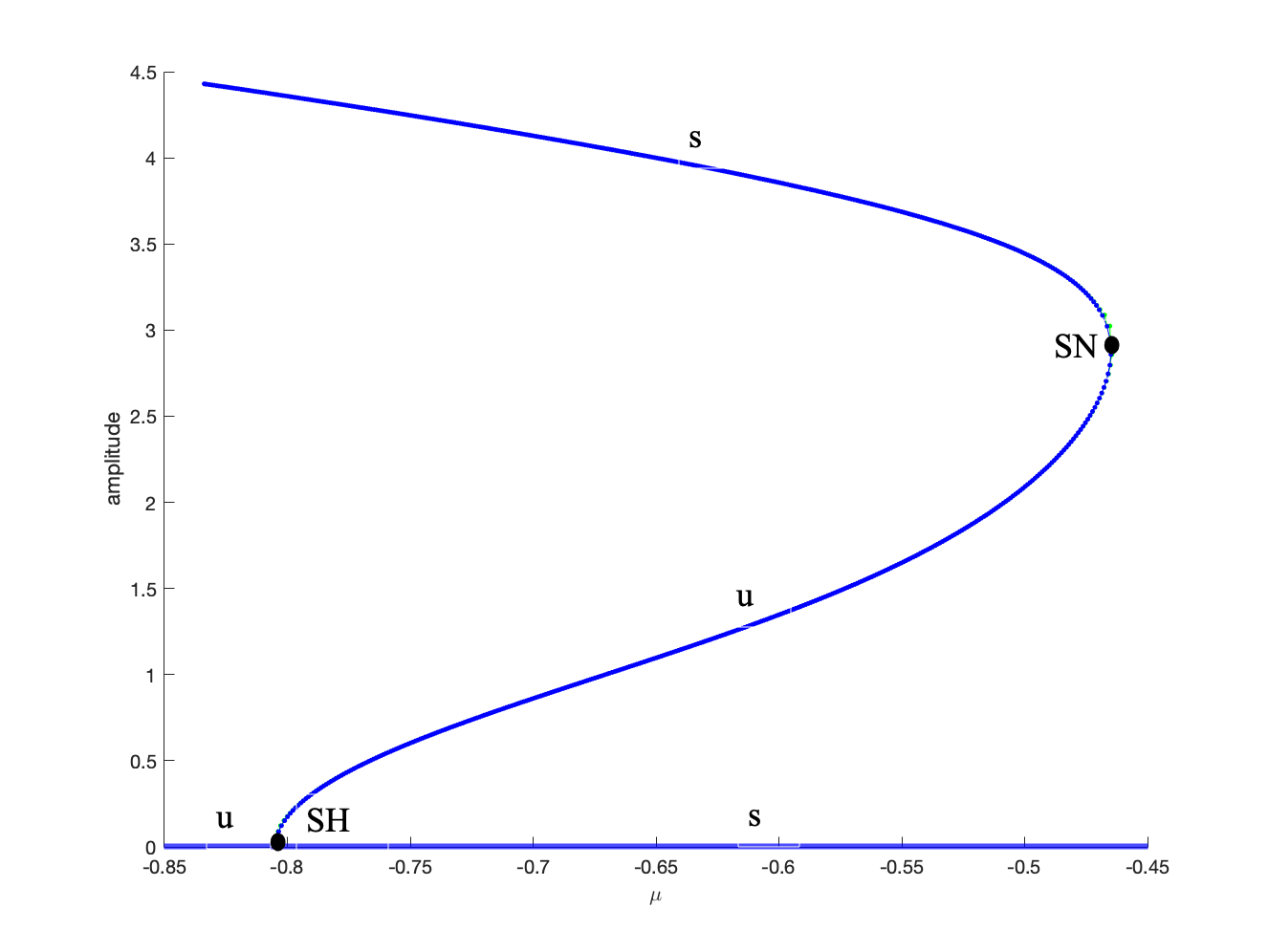}
    \caption{Branches of periodic solutions and equilibrium points from numerical continuation with respect to $\mu$. A hysteresis bifurcation is induced by a subcritical Hopf bifurcation (SH) at $\mu_c = -0.8048$ and a saddle-node bifurcation of limit cycles (SN) at $\mu_f = -0.4649$. Parameter values are $a=0.7$, $b=0.8$, $\rho = 0.5$ and $\tau = 15$. The symbol 's' denotes stable and 'u' denotes unstable.}
    \label{fig:ddebiftool-fig4}
\end{figure}

The system with large time delay generally has several Hopf bifurcations over short parameter intervals \cite{Wolfrum:2010}. It also occurs in our system. For the numerical analysis, we choose the Hopf point $\mu_c = -0.8048$, where the equilibrium point becomes stable as $\mu$ increases and passes through $\mu_c$. The corresponding eigenvalues at the point are $\pm w_ci = \pm 0.6237i$ and the first Lyapunov coefficient is $ L_1 = 0.0310$, which indicates that the system experiences a subcritical Hopf bifurcation at the critical point. In addition, a saddle-node bifurcation of limit cycles occurs at $\mu_f = -0.4649$. The bifurcation diagram carried out with DDE-BIFTOOL is shown in Fig.~\ref{fig:ddebiftool-fig4}.  

\begin{figure}[ht]
     \centering
     \begin{subfigure}[b]{0.6\textwidth}
         \centering
         \includegraphics[width=\textwidth]{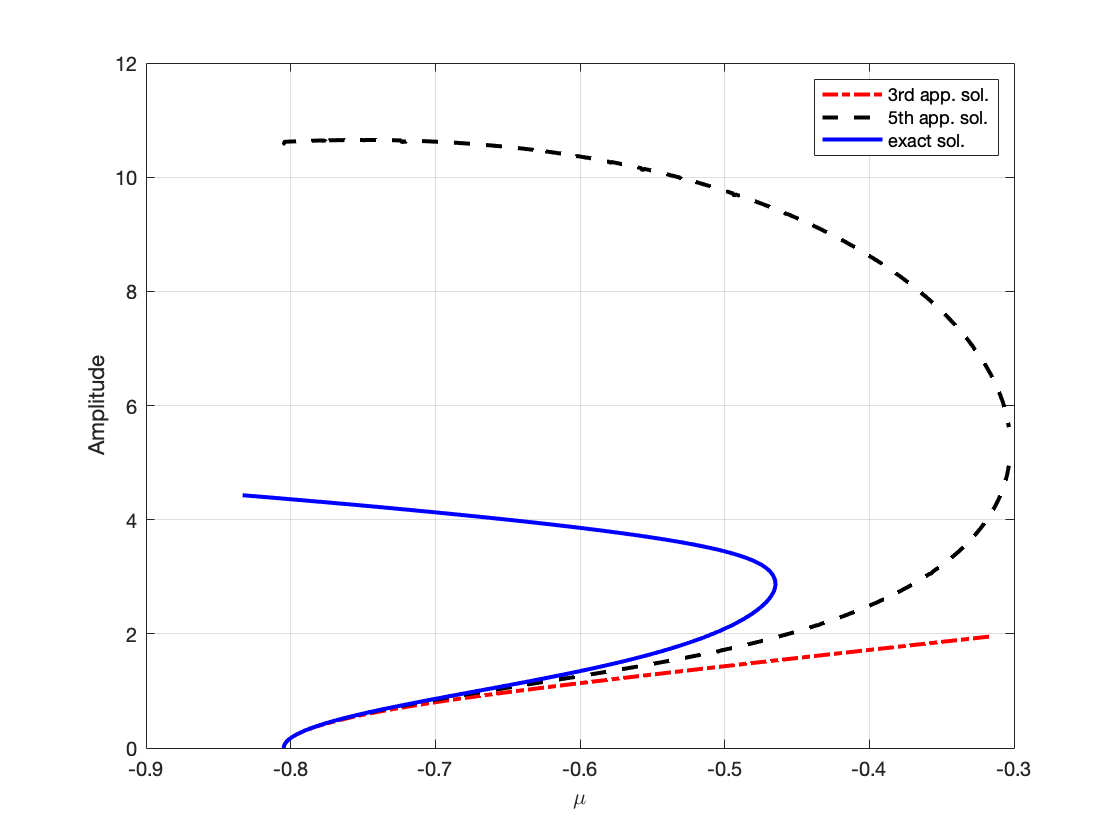}
     \end{subfigure}
     \hfill
     \begin{subfigure}[b]{0.6\textwidth}
         \centering
         \includegraphics[width=\textwidth]{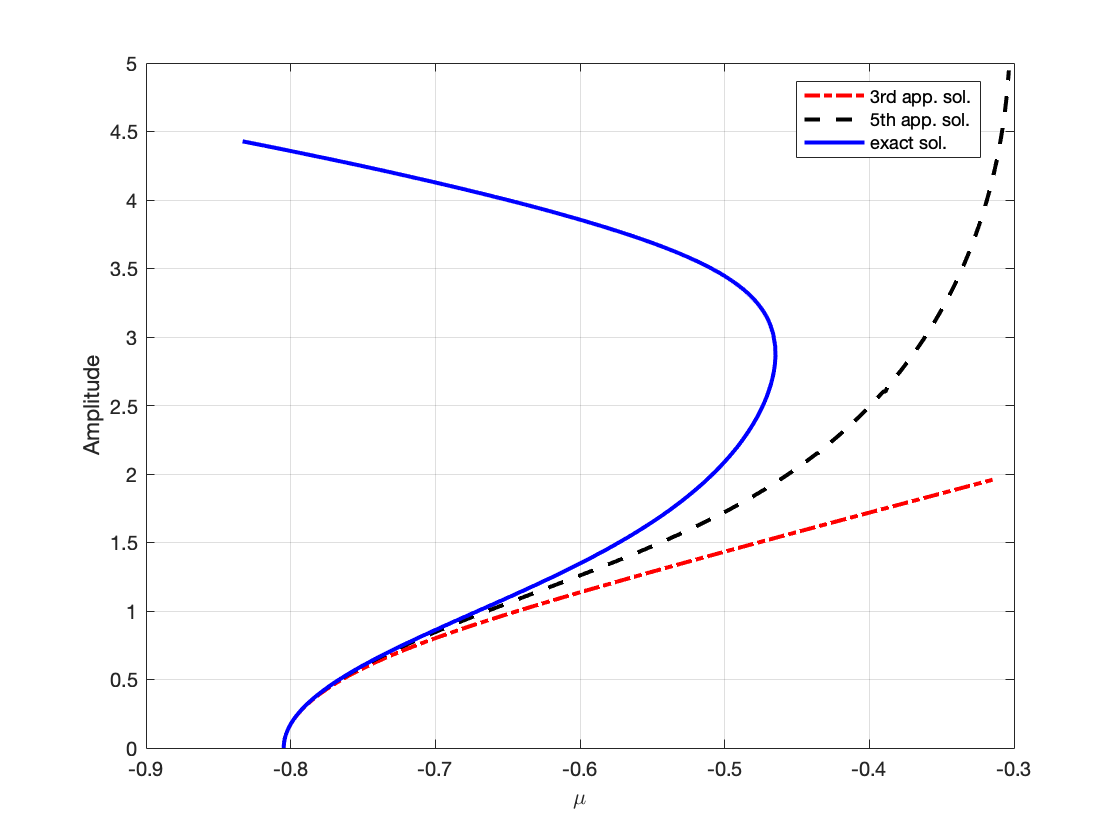}
     \end{subfigure}
     \caption{Bifurcation diagram comparison between the original system carried out by DDE-BIFTOOL (blue solid), the third-order (red dashdot) and the fifth-order (black dash) approximate solutions by the method of multiple scales.  (b) shows details in the neighbourhood of the Hopf point $\mu_c = -0.8048$.}
     \label{fig:Comp_AppExact}
\end{figure}

On the other hand, we found the Landau coefficients of the 3rd-order normal form from (\ref{eq-FN-SLE3-coefficient}): when $\mu < \mu_c$, $\alpha_3 = 0.0738-0.0040i$ and $\beta_3 = 0.0043-0.0023i$; when $\mu > \mu_c$, $\alpha_3$ change the sign and becomes $-0.0738+0.0040i$, whereas $\beta_3$ keep the same. Based on (\ref{eq-SLE3-solutions}) and (\ref{eq-SLE3-eigenvalues}), we can see that in the vicinity of the critical point $\mu_c=-0.8048$, the system exhibits a subcritical Hopf bifurcation, which is consistent with the numerical results of DDE-BIFTOOL. In addition, we have the Landau coefficients of the 5th-order normal form from (\ref{eq-FN-SLE5-coefficient}):  when 
$\mu < \mu_c$: $\alpha_5 = -0.0810+0.0088i$, $\beta_5 = -0.0026-0.0015i$ and $c_5 = -1.3765 \times 10^{-4} - 1.5331 \times 10^{-4}i$. When $\mu > \mu_c$, $\alpha_5$ and $c_5$ remain the same, whereas $\beta_5$ changes sign across the Hopf point. We choose $\varepsilon = |\mu_c-\mu_f|/5 = 0.068 \ll 1$ in (\ref{eq-FN-SLE51}) to meet the requirement of the parameter expansion in (\ref{eq-bifur-mu}).  By analysis of  the solutions of the normal form (\ref{FN-SLE52a}) using (\ref{eq-SLE5-solution}) and (\ref{eq-SLE5-Jacobi}), we obtain the following stability results: 
\begin{align*}
  \mu < \mu_c: \quad & J(r_1)= 0.073 > 0, &  J(r_2) &= -31.21 < 0;  \\
  \mu > \mu_c: \quad & J(r_1)=-0.074 < 0, &  J(r_2) &= -30.63 <0,  \quad  J(r_3) = 0.15 > 0.
\end{align*}
This shows that before the Hopf point $\mu_c$, there are one unstable equilibrium point and one stable periodic orbit; after $\mu_c$, the equilibrium point becomes stable and there exist two periodic solutions, one is stable and the other is unstable. The results are consistent with  the numerical continuation in Fig.~\ref{fig:ddebiftool-fig4}, where a hysteretic loop is formed between  $\mu_c = -0.8048$ and $\mu_f = -0.4649$.
To further prove precision of our results, the bifurcation diagram comparison is shown in  Fig.~\ref{fig:Comp_AppExact} between the original delayed FHN system carried out with DDE-BIFTOOL and the analytical results derived by the method of multiple scales. One can see that close to the Hopf point $\mu_c = -0.8048$, the approximate solutions expanded to third order and fifth order are in good agreement with the exact solution. As expected, the fifth order expansion is better. We also notice that in the region far away from the Hopf point, that is, where $\varepsilon \ll 1$ is not satisfied, great detachment occurs. This is the restriction of the weakly nonlinear analysis. However, compared with expansion to third order, the fifth-order solution is capable enough to predict the existence of the other limit cycle.

\begin{figure}[htbp]
  \centering
  \includegraphics[width=.9\linewidth]{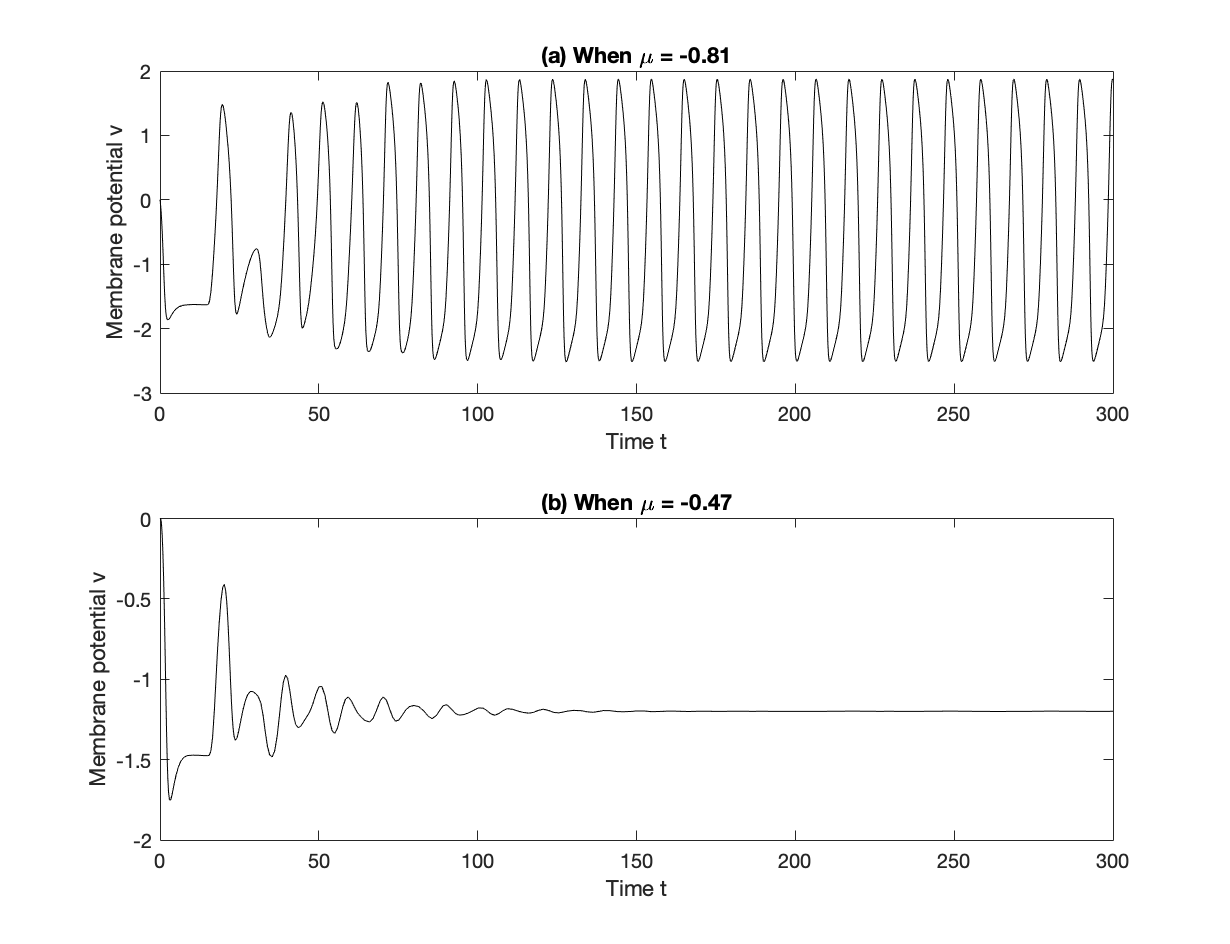}  
  \caption{Numerical simulations showing stable behaviours for (a) $\mu = -0.81$ and (b) $\mu = -0.47$. The initial conditions are $v(t) = 0$ and $w(t) = w_0$. Other parameter values are as in Fig.~\ref{fig:ddebiftool-fig4}.}
\label{fig:dde23-stable}
\end{figure}

\begin{figure}[htbp]
    \centering
    \includegraphics[width=0.9\textwidth]{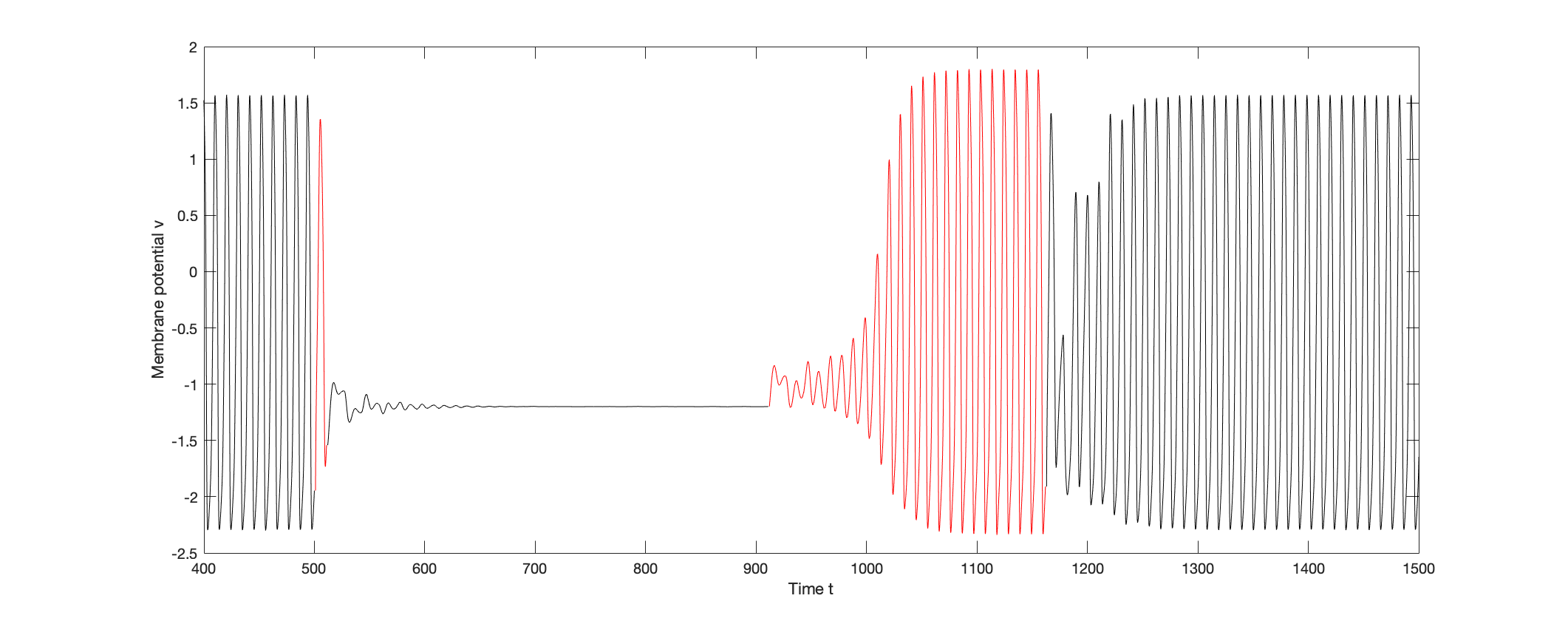}
    \caption{Numerical simulations showing bistability between the equilibrium point solution and periodic orbit solution. Parameter values are as in Fig.~ \ref{fig:ddebiftool-fig4} and $\mu = -0.6$. Switchings between the attractors are achieved by applying two perturbations to the parameter $a$ as follows: $\Delta a = 0.01$, $501\leq t \leq 511$; $\Delta a = -0.2$, $912 \leq t \leq 1162$.}
    \label{fig:dde23-bistable}
\end{figure}

In addition, Fig.~\ref{fig:dde23-stable} shows the numerical solutions for $\mu = -0.81$ and for $\mu = -0.47$, respectively to show stable behaviours beyond the hysteresis region. Fig.~\ref{fig:dde23-bistable} gives numerical simulations to demonstrate bistability for $\mu = -0.6$, where the system switches between the periodic orbit and equilibrium point in the hysteresis region. It seems that the basin of attraction of the equilibrium point is larger than that of the limit cycle, such that a perturbation with a longer term and greater strength is required to drive the system to escape from the equilibrium point and approach the limit cycle.

\section{Conclusions}
Over the years, there have been a substantial number of purely experimental work with phenomenological descriptions of the remarkable dynamical behaviour: hysteresis. A mathematical appreciation of such dynamics must deal with the analysis from the dynamical system point of view by using bifurcation and perturbation theories. In this paper, we have summarized some types of hysteresis bifurcations and shown biological examples to illustrate these phenomena. We have classified hysteresis in terms of catastrophic transitions between different types of attractors. Hysteretic dynamics can be easily appreciated when only involving equilibrium points. Situations become complicated when involving cycles, multiple attractors and/or complex, even global bifurcations. Correspondingly, the theoretical analysis becomes more difficult.

We have theoretically investigated the instance where hysteretic movements between the equilibrium point and the limit cycle are initiated  by a subcritical Hopf bifurcation and a saddle-node bifurcation of limit cycles. We have applied the method of multiple scales in the time-delayed FitzHugh-Nagumo neural system close to the Hopf point and reduced the  governing equations to a fifth-order normal form without delays. From the normal form, we can predict the amplitude and frequency of stable and unstable limit cycles, and the region of hysteresis with bistability. Before the expansion, we need information about the value of the bifurcation parameter at the Hopf point, the marginally stable eigenvalues and the corresponding direct and adjoint eigenvectors. The later process of analytical expansion may be lengthy and tedious, but the procedure is standard and can be automatically realized by symbolic solvers, such as Maple \cite{SANCHEZ:1996}. Our theoretical results have shown good agreement with numerical simulations and continuation.

In addition, we should point out that the normal form derived from the parameter expansion is strictly valid only for the vicinity of the Hopf point, where $\varepsilon \ll 1$ in (\ref{eq-bifur-mu}). If the saddle-node point doesn't fall in this region, expansion to a higher-order normal form is required to be in agreement with the numerical continuation. However, the fifth-order normal form is capable enough to predict the existence of the stable limit cycle,  another attractor required in the hysteresis region. In fact, a similar equation to (\ref{FN-SLE52a}) is often used to illustrate a saddle-node bifurcation of limit cycles in the literature. And such a bifurcation, including its counterpart involving equilibrium points, frequently appear in forming a hysteretic loop. Finally, we have  performed the analysis on a specific system, but the ideas and relevant procedures can be generalized to other systems or other bifurcation parameters, especially the time delay, to investigate the unignorable influence on system dynamics. Since the study of bursting oscillations in neuroscience can often be put in the general framework of hysteresis dynamics, our approach may be useful in analyzing bursting dynamics.

\section*{Acknowledgments}
This work benefited from the support of the Natural Science and Engineering Research Council of Canada.

\appendix
\section{Expressions when solving the differential equations (\ref{eq-FN-MMS2})-(\ref{eq-FN-MMS5}) }
\begin{enumerate}
    \item Expressions of (\ref{eq-FN-MMS-u2})
\begin{subequations}\label{eq-FN-MMS-u2-2}
  \begin{align}
    \boldsymbol{U}_2^{|W|^2} 
    & =  
    M_0^{-1} \boldsymbol{F}_2^{|W|^2}  
    \equiv  
    \left (
    \begin{array}{c}
        X_2^{|W|^2} \\
        Y_2^{|W|^2} \\
    \end{array}
    \right ), \\
    \boldsymbol{U}_2^{W^2}  & = M_{2iw_c}^{-1} \boldsymbol{F}_2^{W^2}  \equiv  \left (
                      \begin{array}{c}
                        X_2^{W^2} \\
                        Y_2^{W^2} \\
                      \end{array}
                    \right ) ,
\end{align}
\end{subequations}
where the non-singular matrices $M_0$ and $M_{2iw_c}$ are from (\ref{eq-FN-MMS-eigenvalue}) with $s=0$ and $s = 2iw_c$, respectively.

    \item Expressions of (\ref{eq-FN-MMS-u3})
\begin{subequations}\label{eq-FN-MMS-u3-2}
  \begin{align}
    \boldsymbol{U}_3^W             & =  M_{iw_c}^{+1}\big(\boldsymbol{F}_3^W-\alpha_3H\boldsymbol{q}\big),  \\
    \boldsymbol{U}_3^{|W|^2W}      & = M_{iw_c}^{+1}\big(\boldsymbol{F}_3^{|W|^2W}-\beta_3H\boldsymbol{q}\big), \\
    \boldsymbol{U}_3^{W^3}        & =  M_{3iw_c}^{-1}\boldsymbol{F}_3^{W^3},
\end{align}
\end{subequations}
where $M^{+1}$ means pseudo inverse of the matrix $M$ because the matrix $M_{iw_c}$ is singular.

\begin{subequations}\label{eq-FN-MMS-F3W}
\begin{align}
  \boldsymbol{F}_3^W & = \left(
            \begin{array}{c}
              \delta_2X_1^We^{-iw_c\tau} \\
              0 \\
            \end{array}
          \right),  \\
  \boldsymbol{F}_3^{|W|^2W} & = \left(
                        \begin{array}{c}
                          -2v_0\Big( X_1^WX_2^{|W|^2}+\overline{X_1^W}X_2^{W^2} \Big) -X_1^W|X_1^W|^2 \\
                          0 \\
                        \end{array}
                      \right), \\
  \boldsymbol{F}_3^{W^3} & = \left(
              \begin{array}{c}
                -2v_0X_1^WX_2^{W^2}-\frac{1}{3}\big( X_1^W\big)^3 \\
                0 \\
              \end{array}
            \right).
\end{align}
\end{subequations}
Here, $\overline X$ means complex conjugate.

   \item Expressions of (\ref{eq-FN-MMS-u4})
   
\begin{subequations}
\begin{align}
   \begin{split}
    \boldsymbol{F}_4^{|W|^4} 
    & =
    -2(I+\tau\mu_cB){\rm Re}(\beta_3)\boldsymbol{U}_2^{|W|^2} \\
    & \qquad
    -
    \left(
    \begin{array}{c}
        v_0 \left ( X_2^{|W|^2}\right)^2 
        + 
        2v_0 \left| X_2^{W^2}\right|^2  \\
        0 \\
    \end{array}
    \right) 
    -
    \left(
    \begin{array}{c}
         2\left| X_1^W\right|^2X_2^{|W|^2} \\
        0 \\
    \end{array}
    \right) \\
    & \qquad
    +
    \Bigg [
    \left(
    \begin{array}{c}
        -2v_0\overline{X_1^W}X_3^{|W|^2W} \\
        0 \\
    \end{array}
    \right) 
    +
    \left(
    \begin{array}{c}
       -\left( X_1^W\right)^2 \overline{X_2^{W^2}}  \\
       0 \\
    \end{array}
    \right)
    + c.c.
    \Bigg ]
  \end{split}
  \\
  \begin{split}
    \boldsymbol{F}_4^{|W|^2} 
    &= 
    -2(I+\tau\mu_cB){\rm Re}(\alpha_3)\boldsymbol{U}_2^{|W|^2} 
    +
    \delta_2 B \boldsymbol{U}_2^{|W|^2} \\
    & \qquad
    +
    \Bigg [
    \left(
    \begin{array}{c}
        -2v_0\overline{X_1^W}X_3^W  \\
        0 \\
    \end{array}
    \right)
    +c.c.
    \Bigg ]
  \end{split}
  \\
  \begin{split}
    \boldsymbol{F}_4^{W^2} 
    &=
    -2(I+\tau\mu_cBe^{-2iw_c\tau})\alpha_3 \boldsymbol{U}_2^{W^2}
    +
    \delta_2 B \boldsymbol{U}_2^{W^2} e^{-2i w_c \tau} 
    \\
    & \qquad
    - 
    \left(
    \begin{array}{c}
       2v_0X_1^WX_3^W \\
       0 \\
    \end{array}
    \right)
  \end{split}
  \\
  \begin{split}\label{eq-FN-MMS-F4W42}
    \boldsymbol{F}_4^{|W|^2W^2} 
    &=
    -2(I+\tau\mu_cBe^{-2iw_c\tau})\beta_3 \boldsymbol{U}_2^{W^2}
    -
    \left(
    \begin{array}{c}
        2v_0X_2^{|W|^2}X_2^{W^2}\\
        0 \\
    \end{array}
    \right) \\
    & \qquad
    - 
    \left(
    \begin{array}{c}
        2v_0X_1^WX_3^{|W|^2W}+2v_0\overline{X_1^W}X_3^{W^3} \\
        0 \\
    \end{array}
    \right) \\
    & \qquad
    -
    \left(
    \begin{array}{c}
        X_2^{|W|^2}\left(X_1^W\right)^2 + 2\left|X_1^W\right|^2X_2^{W^2} \\
        0 \\
    \end{array}
    \right)
  \end{split}
  \\
  \begin{split}\label{eq-FN-MMS-F4W43}
    \boldsymbol{F}_4^{W^4} 
    &= 
    \left(
    \begin{array}{c}
        -v_0\left( X_2^{W^2}\right)^2 -2v_0X_1^WX_3^{W^3}-\left( X_1^W\right)^2X_2^{W^2}\\
        0 \\
    \end{array}
    \right)
  \end{split}
\end{align}
\end{subequations}


\begin{subequations}\label{eq-FN-MMS-u4-2}
  \begin{align}
    \boldsymbol{U}_4^{|W|^2}       & = M_0^{-1}\boldsymbol{F}_4^{|W|^2},  \\
    \boldsymbol{U}_4^{|W|^4}       & = M_0^{-1}\boldsymbol{F}_4^{|W|^4}, \\
    \boldsymbol{U}_4^{W^2}         & = M_{2iw_c}^{-1}\boldsymbol{F}_4^{W^2},  \\
    \boldsymbol{U}_4^{|W|^2W^2}    & = M_{2iw_c}^{-1}\boldsymbol{F}_4^{|W|^2W^2}, \\
    \boldsymbol{U}_4^{W^4}         & = M_{4iw_c}^{-1}\boldsymbol{F}_4^{W^4}.
\end{align}
\end{subequations}

   \item Expressions of (\ref{eq-FN-MMS51})
\begin{subequations}
\begin{align}
   \begin{split}
     \boldsymbol{F}_5^W 
    & =
    -\left(
    I+\tau\mu_cBe^{-iw_c\tau}
    \right)\alpha_3 \boldsymbol{U}_3^W \\
    & \qquad 
    +
   \delta_2Be^{-iw_c\tau}(\boldsymbol{U}_3^W -\alpha_3\tau \boldsymbol{U}_1^W)   
   \end{split}     
   \\
  \begin{split}
    \boldsymbol{F}_5^{|W|^2W} 
  & =
  -\left(
  I+\tau\mu_cBe^{-iw_c\tau}
  \right)
  \left(
  \beta_3 \boldsymbol{U}_3^W 
  +
  \big(
  2\alpha_3+\overline{\alpha_3}
  \big)
  \boldsymbol{U}_3^{|W|^2W}
  \right) \\
  & \qquad 
  +
  \delta_2 B \boldsymbol{U}_3^{|W|^2W} e^{-iw_c\tau} 
  - \delta_2B\tau \beta_3 \boldsymbol{U}_1^W e^{-iw_c\tau} \\
  & \qquad 
  -2v_0 
  \left(
  \begin{array}{c}
    X_2^{|W|^2}X_3^W+X_2^{W^2}\overline{X_3^W}+X_4^{|W|^2}X_1^W+\overline{X_1^W}X_4^{W^2} \\
    0 \\
  \end{array}
  \right) \\
  & \qquad 
  - \left(
  \begin{array}{c}
    2|X_1^W|^2X_3^W+(X_1^W)^2\overline{X_3^W} \\
    0 \\
  \end{array}
  \right) 
  \end{split}  
  \\
  \begin{split}
  \boldsymbol{F}_5^{|W|^4W} 
  &=
  -\left(
  I+\tau\mu_cBe^{-iw_c\tau}
  \right)
  \big(
  2\beta_3+ \overline{\beta_3}
  \big)
  \boldsymbol{U}_3^{|W|^2W} \\
  & \qquad
  -2v_0
  \left(
    \begin{array}{c}
       X_2^{|W|^2}X_3^{|W|^2W}+X_2^{W^2}\overline{X_3^{|W|^2W}}+\overline{X_2^{W^2}}X_3^{W^3} \\
       0 \\
    \end{array}
  \right) \\
  & \qquad
  -2v_0
  \left(
    \begin{array}{c}
       X_4^{|W|^4}X_1^W+\overline{X_1^W}X_4^{|W|^2W^2} \\
       0 \\
    \end{array}
   \right) \\
   & \qquad
  -\left(
     \begin{array}{c}
        2|X_1^W|^2 X_3^{|W|^2W} 
        +
        (X_1^W)^2 \overline{X_3^{|W|^2W}} 
        +
        \overline{(X_1^W)^2} X_3^{W^3} \\
        0 \\
     \end{array}
   \right) \\
   & \qquad
   -\left(
     \begin{array}{c}
        X_1^W \Big( X_2^{|W|^2} \Big)^2 
        + 
        2\overline{X_1^W} X_2^{|W|^2} X_2^{W^2}
        +
        2X_1^W |X_2^{W^2}|^2 \\
        0 \\
     \end{array}
   \right)
\end{split}
\\
\begin{split}
   \boldsymbol{F}_5^{|W|^2W^3} 
  &= 
  -(I + \tau \mu_c B e^{-3i w_c \tau}) 3\beta_3 \boldsymbol{U}_3^{W^3} \\
  & \qquad
  -2v_0
  \left(
  \begin{array}{c}
     X_2^{|W|^2} X_3^{W^3} 
     +
     X_2^{W^2} X_3^{|W|^2W}     \\
     0 
  \end{array}
  \right) \\
  & \qquad
  -2v_0
  \left(
  \begin{array}{c}
     X_1^W X_4^{|W|^2W^2}
     +
     \overline{X_1^W} X_4^{W^4}     \\
     0 
  \end{array}
  \right) \\
  & \qquad
  -\left(
  \begin{array}{c}
     2|X_1^W|^2 X_3^{W^3} 
      +
     (X_1^W)^2 X_3^{|W|^2W}    \\
     0 
  \end{array}
  \right) \\
  & \qquad
  -\left(
  \begin{array}{c}
     \overline{X_1^W} \Big( X_2^{W^2} \Big)^2
      +
     2X_1^W X_2^{|W|^2} X_2^{W^2}    \\
     0 
  \end{array}
  \right)
\end{split}
\\
\begin{split}
    \boldsymbol{F}_5^{W^3}
    &=
    -(I + \tau \mu_c B e^{-3i w_c \tau}) 3\alpha_3 \boldsymbol{U}_3^{W^3} 
    + \delta_2 B \boldsymbol{U}_3^{W^3} e^{-3i w_c \tau} \\
    & \qquad
    -2v_0
    \left(
    \begin{array}{c}
         X_2^{W^2} X_3^W  \\
         0 
    \end{array}
    \right)
    -2v_0
    \left(
    \begin{array}{c}
         X_1^{W} X_4^{W^2}  \\
         0 
    \end{array}
    \right) \\
    & \qquad
    -\left(
    \begin{array}{c}
        (X_1^W)^2 X_3^W   \\
         0 
    \end{array}
    \right)
\end{split}
\\
\begin{split}
    \boldsymbol{F}_5^{W^5}
    &=
    -2v_0  
    \left(
    \begin{array}{c}
         X_2^{W^2} X_3^{W^3}  \\
         0 
    \end{array}
    \right)
    -2v_0  
    \left(
    \begin{array}{c}
         X_1^W X_4^{W^4}   \\
         0 
    \end{array}
    \right) \\
    & \qquad
    -  
    \left(
    \begin{array}{c}
         (X_1^W)^2 X_3^{W^3}   \\
         0 
    \end{array}
    \right)
    -  
    \left(
    \begin{array}{c}
         X_1^W \Big( X_2^{W^2} \Big)^2   \\
         0 
    \end{array}
    \right)
\end{split}
\end{align}
\end{subequations}

\begin{subequations}\label{eq-FN-MMS-u5-2}
\begin{align}
    \boldsymbol{U}_5^W & = M_{iw_c}^{+1} 
    \left ( 
    \boldsymbol{F}_5^W - \alpha_5 H q
    \right ), \\
    \boldsymbol{U}_5^{|W|^2W} & = M_{iw_c}^{+1} 
    \left (
    \boldsymbol{F}_5^{|W|^2W} - \beta_5 H q
    \right ), \\
    \boldsymbol{U}_5^{|W|^4W} & = M_{iw_c}^{+1}
    \left (
    \boldsymbol{F}_5^{|W|^4W} - c_5 H q
    \right ), \\
    \boldsymbol{U}_5^{|W|^2 W^3} & = M_{3iw_c}^{-1} \boldsymbol{F}_5^{|W|^2 W^3}, \\
    \boldsymbol{U}_5^{W^3} & = M_{3iw_c}^{-1} \boldsymbol{F}_5^{W^3}, \\
    \boldsymbol{U}_5^{W^5} & = M_{5iw_c}^{-1} \boldsymbol{F}_5^{W^5}.
\end{align}
\end{subequations}

\end{enumerate}


\bibliography{mybib}

\end{document}